\documentclass[mathleft]{an}
\usepackage{graphicx}
\usepackage{times}
\usepackage{natbib}
\begin{document}


\title{P-modes in rapidly rotating stars -- \\
        looking for regular patterns in synthetic asymptotic spectra}

\author{F. Ligni\`eres \inst{1,2}\fnmsep\thanks{Corresponding author:
  \email{francois.lignieres@ast.obs-mip.fr}\newline}
\and B. Georgeot\inst{3,4}
\and J. Ballot\inst{1,2}
}
\titlerunning{P-modes in rapidly rotating stars}
\authorrunning{F. Ligni\`eres, B. Georgeot \& J. Ballot}
\institute{
Universit\'e de Toulouse; UPS; Laboratoire d'Astrophysique de Toulouse-Tarbes (LATT); F-31400 Toulouse, France
\and
CNRS; Laboratoire d'Astrophysique de Toulouse-Tarbes (LATT); F-31400 Toulouse, France
\and
Universit\'e de Toulouse; UPS; Laboratoire de
Physique Th\'eorique (IRSAMC); F-31062 Toulouse, France
\and
CNRS; LPT (IRSAMC); F-31062 Toulouse, France}



\abstract{%
According to a recent ray-based asymptotic theory, the high-frequency
p-mode spectrum of rapidly rotating stars is a superposition of frequency subsets associated 
with dynamically independent regions of the ray-dynamics phase space.
At high rotation rates corresponding to typical $\delta$ Scuti stars, two frequency subsets are expected to be visible :
a regular frequency subset described by a Tassoul like formula and an irregular frequency subset with specific statistical properties. 
In this paper, we investigate whether the regular patterns can be detected in the resulting spectrum.
We compute the autocorrelation function of synthetic spectra where the frequencies follow the asymptotic theory, 
the relative amplitudes are simply given by the modes' disk-averaging factors,
and the frequency resolution is that of a CoRoT long run.
Our first results are that (i) the detection of regular patterns strongly depends on the ratio of regular over irregular modes,
(ii) low inclination angle configurations are more favorable than near equator-on configurations
(iii) in the absence of differential rotation, the $2 \Omega$ rotational splitting between $m=1$ and $m=-1$
modes is an easy feature to detect.}

\maketitle

\section{Introduction}

The launching of the space missions COROT and KEPLER
is bringing a new wealth of observational data on the
asteroseismology of many different types of stars.
For slowly rotating stars, e.g. the sun, the approximate spherical
symmetry of the system enables us to classify modes according
to well-defined sets of integers. This mode structure 
allowed us to extract information on the stellar interior from
the spectrum of oscillation modes \citep{obssun}.

However, for rapidly rotating stars, the deformation of the star
breaks the spherical symmetry.  It has recently been shown
that perturbation theory fails to yield accurate predictions 
of oscillation spectra even for stars with moderate rotation rates \citep{Lign,Rees}. 
In \cite{LG1,LG2}, an asymptotic theory of spectra of such stars
was built, based on acoustic ray dynamics. It was shown that this dynamics
is Hamiltonian, and undergoes
a gradual transition from integrability to chaos when the rotation increases.
At moderately rapid rotation, the phase space is divided
into integrable and chaotic zones, where the
dynamics is qualitatively different.
It was shown in \cite{LG1,LG2} that this structure modifies the structure
of the oscillation spectra.  The acoustic mode frequencies
cannot be in general associated to well-defined sets of integers.
Instead, the spectrum is split into independent subspectra,
corresponding to different phase space zones for the ray dynamics.
The subspectra corresponding to integrable zones give rise to
regular sequences of frequencies, whereas a superimposed subspectrum
is associated to chaotic dynamics and displays regularity only in a statistical
sense.

In order to connect such results to observed spectra, several
questions have to be explored.  The first one is the validity of the asymptotic
theory to the finite range of spectra which can be observed. A first answer
was given in \linebreak \cite{LG1,LG2}, where it was shown that the predicted
structure can be found in numerically \linebreak computed
relatively low-frequency p-modes of polytropic stellar models. An other important question
is to separate these different subspectra in observational data, where
the visibility of the modes plays an important role. A first step
in this direction is to construct synthetic spectra based
on the asymptotic theory where visibility varies
from mode to mode and try to extract from
them important features of the system, as if they were observed spectra.
This is the strategy we follow here.
We also note that a first evidence of regular patterns
in a p-mode spectrum of a rapidly rotating star has been recently obtained
by analyzing the spectra of a $\delta$ Scuti CoRoT target \citep{Hern}.

\section{Construction of synthetic spectra in the asymptotic regime}

We have built synthetic spectra with the following formulas:
\begin{equation}
 S(\nu, i) = \sum_j L(\nu, \nu_j, A_j(i), \Gamma)	
\label{synth}
\end{equation}
\begin{equation}
L(\nu, \nu_j, A_j(i), \Gamma) = A_j(i) \exp \left[ -(\nu-\nu_j)^2/\Gamma^2 \right]
\end{equation}
\noindent
where $\nu_j$ are the frequencies of the modes and $A_j(i)$ their 
amplitudes, which are functions of the inclination angle $i$. We have 
represented modes with Gaussian functions of \linebreak width $\Gamma$ to take into 
account the finite resolution of the spectrum. We have considered in 
this work a resolution $\Gamma$ corresponding to 150-day-long runs.
Our main assumptions are that the frequencies follow the asymptotic behaviour described in \cite{LG2}
and that the relative amplitudes only depend on the relative mode visibilitites, the visibility being 
approximated by the disk-averaging factor computed for high-frequency p-modes in a polytropic stellar model (see formula (37) in \cite{LG2}).
Above a certain rotation rate, the p-modes spectrum is dominated by two families of modes, the island \linebreak modes and the chaotic modes, as the visibilities of the other modes become negligible.
The island modes are associated with island chain structures of the ray dynamics phase space and their spectrum
follows a simple formula \citep{Rees8}.
\begin{equation}
\omega_{\tilde{n}\tilde{\ell} m} = \Delta_{\tilde{n}} {\tilde{n}} + \Delta_{\tilde{{\ell}}} {\tilde{{\ell}}} + \Delta_{m} | m | -m \Omega + \omega_{\mathrm{ref}}
\label{asymp}
\end{equation}
\noindent
where $\tilde{\ell} \le 3$ and $\Delta_{\tilde{n}}, \Delta_{\tilde{{\ell}}}, \Delta_{m}$ depend on the star's structure.
The quantum numbers $\tilde{n}$ and $\tilde{\ell}$ are defined from the spatial distribution of the modes in a meridional plane. 
They correspond respectively to the number of
nodes along and across the stable periodic trajectory associated with the island chain.
By contrast, chaotic modes are associated with chaotic regions of phase space, 
and their frequency spectrum  is said to be irregular because it is not described by a smooth function
of 3 integers. The frequency spectrum has nevertheless specific statistical properties. 
Indeed, if we consider chaotic modes of the same symmetry class, the distribution of the consecutive frequency spacings
$\sigma_i = \omega_{i+1} - \omega_i$ (scaled by the mean frequency spacing $<\omega_{i+1} - \omega_i>$) is close to the parameter-free Wigner
distribution $P(\sigma) = \pi \sigma / 2 \exp(-\pi \sigma^2/4)$ in accordance with the prediction of Random Matrix Theory.
A symmetry class corresponds to a given azimuthal number $m$ and a given symmetry ($+$ or $-$) with respect to the equator.
For each symmetry class, a spectrum is thus determined from a realization of the Wigner distribution and
the chaotic spectrum is the superposition of all these spectra.
Finally in order to construct the total spectrum, we need to know the ratio between chaotic and island modes. This ratio
tends to increase with rotation and, for the rotation rate considered in the following $\Omega = 0.6 \Omega_K$ where $\Omega_K = \sqrt{GM/R^3}$, it is close to 3.7 \citep{LG2}. 

\section{Autocorrelation of the synthetic spectra}

The autocorrelation of the synthetic spectra has been investigated, for $\Omega = 0.6 \Omega_K$,
by varying three parameters: the frequency range of the spectrum, the inclination angle of the star,
and  an amplitude threshold that only retains the
the highest amplitude peaks of the spectrum.
The parameter domain that we explored is the following :
The frequency range spans $n_{\delta}$ large separations $\delta_n = 2 \Delta_{\tilde{n}}$, where $n_{\delta}$
has been varied from $1$ to $6$. The amplitude threshold is chosen to keep $N_{\delta}$ frequency peaks 
per large separation interval; $N_{\delta}$ has been varied between 10 and 100.
Once $n_{\delta}$ and $N_{\delta}$ are fixed, 
the total number of frequencies of the spectrum is $n_{\delta} \times N_{\delta}$.
The inclination angle has been varied from 0 to $90^{\circ}$.

We find that the inclination angle plays an important role in the search for regular patterns 
because it affects the relative visibility of the chaotic and island modes. 
Indeed, the disk-averaging factor of the chaotic modes tends
to increase towards equator-on configurations while the disk-averaging factor of island modes tends to decrease.

Let us first consider a high inclination configuration $i=63^{\circ}$.
In figures 1 and 2, the autocorrelation of the whole spectrum (top panel), the chaotic spectrum (middle) and the island
spectrum (bottom) are displayed for $N_{\delta}=57$ and for two different frequency ranges $n_{\delta}=3$ (Fig. 1) and 
$n_{\delta}=6$ (Fig. 2).
The autocorrelation is defined as in statistics: it is computed after substracting
the mean and it is normalized by the variance. The autocorrelation is therefore comprised between -1 and 1.
The vertical scale is different for the autocorrelation of the island spectrum 
because, 
as expected, the Tassoul like formula leads to strong peaks associated with the $\Delta_{\tilde{n}}$ regular
spacing and smaller peaks corresponding to linear combinaisons of 
$\Delta_{\tilde{n}}, \Delta_{\tilde{{\ell}}}, \Delta_{m}$.

No structure is seen in the chaotic mode spectrum except for a peaks at $2 \Omega$ (corresponding to the $2 \Omega /\delta_n = 0.836$ peak of Fig. 1).
This lack of structure can be understood from the fact that the chaotic spectrum is a superposition of the
subspectra
corresponding to the different symmetry classes $m^{\pm}$. These subspectra are statistically independent and,
as a consequence, the consecutive difference statistics of the chaotic spectrum is generally close to a Poisson statistics
thus explaining the featureless autocorrelation. The \linebreak pole-on configuration is an exception to this rule as
only axisymmetric modes $m=0$ are visible in this case and
the superposition of only two independent subspectra does not lead to Poisson statistics. 
The $2 \Omega$ peaks is due to a rotational splitting between $m=1$ and $m=-1$ modes.
In the p-modes asymptotic regime, the Coriolis force has indeed a negligible effect on the frequency because its characteristic time-scale
is much larger then the oscillation time-scale. As shown in \cite{Rees}, this asymptotic property is actually already correct at relatively low frequencies.
Since the centrifugal force does not distinguish between $-m$ and $m$ modes, their frequencies can be considered as degenerate in the rotating frame thus
leading to a $2m \Omega$ splitting in the inertial frame.
In addition, the number of $m=1$ and $m=-1$ modes in the spectrum is relatively high because their disk-averaging annulation
effect 
is not 
strong (and less important than for higher $m$ modes), thus leading to a significative peak at $2 \Omega$ in the autocorrelation function.

As shown in the top panel of Fig. 1, 
signatures of the island mode regular patterns are not apparent in the autocorrelation of the total spectrum.
This is due to the fact 
that the ratio of chaotic modes over island
modes is too high in this case.
Increasing the frequency range to 6 large separations (Fig. 2) enables us to detect a peak at the large separation $\delta_n$ 
(the peak seen at a scaled frequency lag of unity on Fig. 1 or Fig. 2). This is expected as a larger frequency range
leads to the build up of the peaks associated with the periodic features of the
frequency spectrum.
For the present high inclination angle configuration, limiting the analysis to the highest amplitude
peaks does not help because in average the visibility of the island modes
is not higher than the visibility of the chaotic modes.

\begin{figure}
\resizebox{\hsize}{!}{\includegraphics{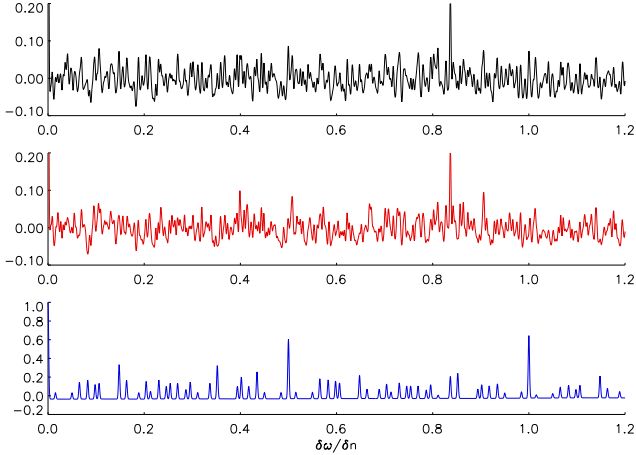}}
\caption{
Autocorrelation of a high-frequency p-modes spectrum resulting from the superposition of a chaotic mode spectrum
and an island mode spectrum. The top panel displays the autocorrelation function of the whole
spectrum, while the middle and bottom panels shows the autocorrelation of the chaotic and island mode spectra, respectively.
The x-axis represents the frequency lag in 
units of the large separation $\delta_n$.
The inclination angle is equal to $i=63^{\circ}$,
the frequency range corresponds to 3 large separations and the $171$ highest amplitude frequencies have been retained in the spectrum.
The only significative feature of the autocorrelation function is the $2 \Omega$ peak also seen in the autocorrelation of the chaotic spectrum.}
\label{fig1}
\end{figure}

\begin{figure}
\resizebox{\hsize}{!}{\includegraphics{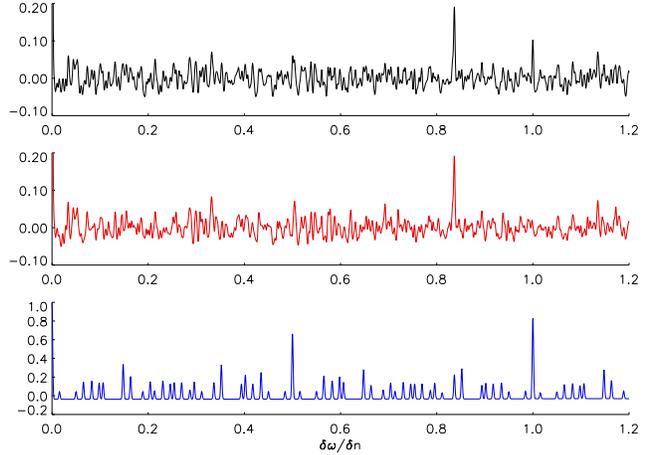}}
\caption{Same as figure 1 except that the frequency range of the spectrum now spans 6 large separations,
the number of frequencies retained per large separation remaining the same.
The autocorrelation spectrum
now shows a peak corresponding to the large separation.}
\label{fig2}
\end{figure}
\begin{figure}
\resizebox{\hsize}{!}{\includegraphics{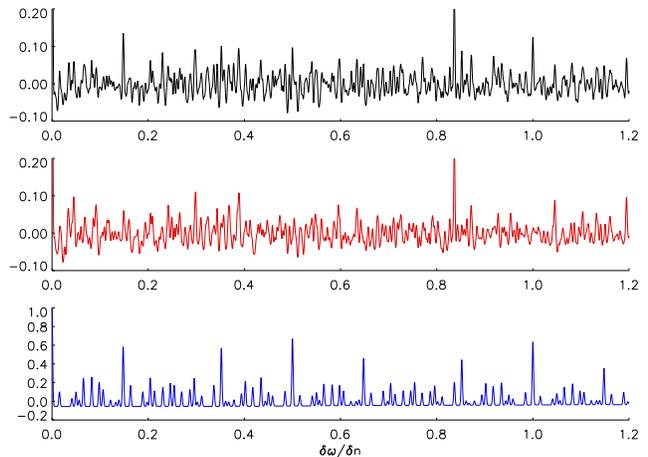}}
\caption{Same as Fig. 1. The inclination angle is now equal to $i=30^{\circ}$,
the frequency range of the spectrum spans 3 large separations and the $204$ highest amplitude frequencies have been retained.
The main correlation peak of the island spectrum can be retrieved in the autocorrelation of the total spectrum.}
\label{fig3}
\end{figure}
\begin{figure}
\resizebox{\hsize}{!}{\includegraphics{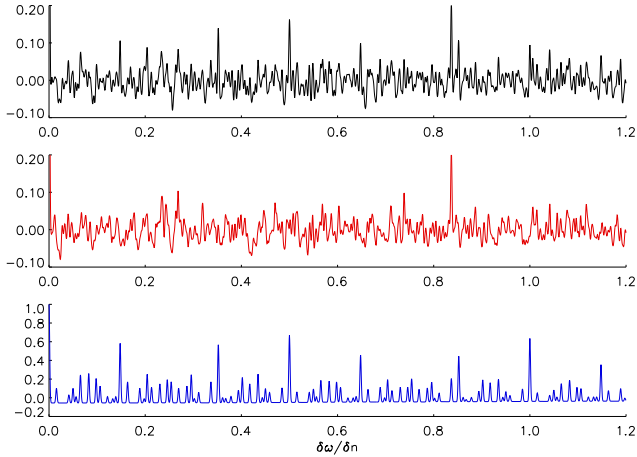}}
\caption{Same as Fig 3. except that
only the $54$ highest amplitude frequencies have been retained. This helps to detect some
of the correlation peaks that were not significant in the previous case.}
\label{fig4}
\end{figure}

As illustrated in figures 3 and 4 for $i=30^{\circ}$, 
low inclination configurations are more favorable in order to detect the island mode regular patterns.
The autocorrelation functions are shown 
for a frequency range $n_{\delta}=3$ and for two different amplitude threshold $N_{\delta}=68$ (Fig. 3) and $N_{\delta}=18$ (Fig. 4).
In both cases, features associated with the island mode regular patterns are detected in the total autocorrelation function.
This is because, as compared to the high inclination case, the proportion of island modes in the total spectrum has increased.
The comparison of figures 3 and 4 also shows that, contrary to the high inclination case, the island mode regular patterns are more easily detected
if the analysis is limited to the highest amplitude modes. 
Finally we checked that, as expected, the signature of these patterns is stronger if the frequency range is increased.

\section{Discussion and conclusion}

Although based on strong simplifying assumptions, the \linebreak present study should provide
some clues to conduct a search for regular patterns in observed
p-mode spectra of rapidly rotating stars.
For example, we find that limiting the search for regular patterns to the highest amplitude
frequency peaks is not necessarily helpful. This depends on the inclination angle.
Generally, low inclination
configurations are more favorable because the island modes are relatively more visible 
than the chaotic modes.
A robust feature of the autocorrelation functions computed in the present study is the $2 \Omega$ peak.
This feature does not rely on the asymptotic regime assumption, it rather requires that the Coriolis force 
has a negligible effect which is already true at relatively low p-mode frequency. 
Possible effects of the
differential rotation on this peak should nevertheless be tested, for example using the formalism described in \cite{Rees09}.

The two most important assumptions in the construction of the synthetic spectra concern the 
asymptotic regime and the mode amplitudes.
We assumed that island mode frequencies strictly follow the asymptotic formula (\ref{asymp}) 
while it is in fact only approximate due to finite wavelength effects
(see the dicussion in \citet{LG2}). The departures from 
the asymptotic formula (\ref{asymp}) have been determined for different stellar models and frequency ranges
in \citet{Lign}, \citet{Rees8} and \citet{Rees09}.
As a first step, a simple way to model these non-asymptotic effects in the present study would be to decrease the frequency resolution
of the synthetic spectra.
A second step would be to compute numerically a complete spectrum using the code TOP \citep{Rees}.
Such computations are nevertheless very time consuming. Thus, simple models of the spectra like
the one presented in this study will still
be useful as they enable us to cover a wider range of parameters.
In what concerns the amplitudes of the modes, 
we lack a 
consistent theory to describe their excitation
and non-linear saturation in rapidly rotating stars.
The present model could nevertheless be 
improved through better determinations of
the  mode visibilities, including 
more realistic stellar models as well
as
non-adiabatic, limb-darkening and gravity darkening effects.
The mode inertia could also be taken into account in modelling the amplitudes.

\acknowledgements
We thank D. Reese for fruitful discussions.
This work
was supported by the Programme National de Physique Stellaire of INSU/CNRS
and the SIROCO project of the Agence National de la Recherche.


\begin{thebibliography}{}
















\bibitem[{Christensen-Dalsgaard(2002)}]{obssun} Christensen-Dalsgaard, J.: 2002, Rev. Mod. Phys. 74, 1073




















                             
\bibitem[Garc{\'{\i}}a Hern{\'a}ndez et 
al.(2009)]{Hern} Garc{\'{\i}}a Hern{\'a}ndez, A., et al.: 2009, A\&A 506, 79

\bibitem[Ligni{\`e}res et al.(2006)]{Lign} Ligni{\`e}res, F.,
Rieutord, M., \& Reese, D.: 2006, A\&A 455, 607

\bibitem[Ligni{\`e}res \& Georgeot(2008)]{LG1} Ligni{\`e}res, F., \& Georgeot, B.: 2008, Pys. Rev. E 78, 016215

\bibitem[Ligni{\`e}res \& Georgeot(2009)]{LG2} Ligni{\`e}res, F., \& Georgeot, B.: 2009, A\&A 500, 1173














\bibitem[Reese et al.(2006)]{Rees} Reese, D., Ligni{\`e}res, F., \& Rieutord, M.: 2006,
A\&A 455, 621


\bibitem[Reese et al.(2008)]{Rees8} Reese, D., Ligni{\`e}res, F., \& Rieutord, M.: 2008,
A\&A 481, 449

\bibitem[Reese et al.(2009)]{Rees09} Reese, D.~R., MacGregor,
K.~B., Jackson, S., Skumanich, A.,
\& Metcalfe, T.~S.: 2009, A\&A 506, 189













\end{thebibliography}
\end{document}